\documentclass[acmsmall]{acmart}
\AtBeginDocument{%
  }
\setcopyright{acmcopyright}
\newcommand{\Space}[1]{}
\usepackage{longtable}
\usepackage{hhline}
\usepackage{booktabs}
\usepackage{hyperref}
\usepackage{color, colortbl}
\usepackage{graphicx}
\definecolor{Gray}{gray}{0.9}

\copyrightyear{2025}
\acmYear{2025}

\begin{document}
\title["Always Nice and Confident, Sometimes Wrong"]{"Always Nice and Confident, Sometimes Wrong": Developer's Experiences Engaging Large Language Models (LLMs) Versus Human-Powered Q\&A Platforms for Coding Support}
\author{Jiachen Li}
\email{li.jiachen4@northeastern.edu}
\affiliation{%
  \institution{Northeastern University}
  \city{Boston}
  \state{Massachusetts}
  \country{USA}
}
\author{Elizabeth D. Mynatt}
\email{e.mynatt@northeastern.edu}
\affiliation{%
  \institution{Northeastern University}
  \city{Boston}
  \state{Massachusetts}
  \country{USA}
}
\author{Varun Mishra}
\email{v.mishra@northeastern.edu}
\affiliation{%
  \institution{Northeastern University}
  \city{Boston}
  \state{Massachusetts}
  \country{USA}
}
\author{Jonathan Bell}
\email{j.bell@northeastern.edu}
\affiliation{%
  \institution{Northeastern University}
  \city{Boston}
  \state{Massachusetts}
  \country{USA}
}
\begin{abstract}
Software engineers have historically relied on human-powered Q\&A platforms like Stack Overflow (SO) as coding aids. With the rise of generative AI, developers have started to adopt AI chatbots, such as ChatGPT, in their software development process. Recognizing the potential parallels between human-powered Q\&A platforms and AI-powered question-based chatbots, we investigate and compare how developers integrate this assistance into their real-world coding experiences by conducting a thematic analysis of 1700+ Reddit posts. Through a comparative study of SO and ChatGPT, we identified each platform’s strengths, use cases, and barriers. Our findings suggest that ChatGPT offers fast, clear, comprehensive responses and fosters a more respectful environment than SO. However, concerns about ChatGPT’s reliability stem from its overly confident tone and the absence of validation mechanisms like SO’s voting system. Based on these findings, we synthesized the design implications for future GenAI code assistants and recommend a workflow leveraging each platform’s unique features to improve developer experiences.

 
\end{abstract}


\maketitle

\section{Introduction}
Software engineers have been adopting many tools to assist their coding process, with online Q\&A platforms standing out as a favored method~\cite{sarka2017knowledge}. Among them, Stack Overflow(SO) is one of the predominant choices for developers to ask programming-related questions~\cite{ahmad2018survey}. Prior research has studied how developers learn and exchange coding knowledge through SO, highlighting its advantages and drawbacks~\cite{treude2011programmers,mamykina2011design,storey2014r}.

With the public release of ChatGPT, a powerful AI chatbot, software engineers swiftly integrated it into their coding practices. This popularity has sparked numerous discussions and trends on social media platforms among developers. For instance, \emph{r/ChatGPTCoding}, focusing on "the coding side of ChatGPT", has over 68,000 members, ranking it top 5\% of all subreddits as of September 2023~\cite{subr2023}. Similar posts can also be found in other subreddits such as \emph{r/learnprogramming} and \emph{r/chatgpt}. These resources on social media offer researchers extensive opportunities to investigate developers' real-world experiences with ChatGPT in assisting their coding practices.

\Space{Realizing the adoption of ChatGPT to assist in coding practices has become increasingly popular, we've noticed a potential shift from traditional online Q\&A platforms to AI-powered chatbots, mainly due to their similarities.} As both operate within a questions and answers interface, ChatGPT has become a potential alternative to platforms like SO~\cite{xue2023can}. However, this transition is not without its challenges. In December 2022, the official SO platform made an announcement prohibiting the use of generative AI (GenAI), including ChatGPT and other Language Model Models (LLMs), for posting content on their platform~\cite{banned2023}. This decision was motivated by concerns that ChatGPT-generated answers could be inaccurate and unreliable, potentially undermining the trustworthiness of the platform~\cite{policy2022}. Given these assessments, comparing human-powered and AI-powered Q\&A platforms has become increasingly important to understand their differences and future potential.

In this research, we aim to investigate how developers incorporate AI-powered Q\&A chatbots into their coding experiences in real-life scenarios and how this experience differs from the traditional practice of posting questions on human-powered Q\&A platforms. To operationalize this goal, we selected two tools as exemplars of these paradigms: ChatGPT and SO. To gain insights into users' real-world experiences, we examined Reddit posts in two categories: those sharing experiences on using ChatGPT for coding assistance and those engaging in discussions specific to comparing ChatGPT with SO, and conducted a thematic analysis to highlight common themes.

Our work focuses on \textbf{two key questions}:
\begin{enumerate}
\item In what ways are individuals employing GenAI chatbots to enhance their coding experience?
\item What differentiates this coding experience from the utilization of conventional human-powered Q\&A platforms?
\end{enumerate}

In answering these questions, we make four key contributions: first, we curated a dataset comprising Reddit posts on ChatGPT for programming after manual filtering. Second, we synthesized insights regarding the \textbf{strengths}, \textbf{use cases}, and \textbf{barriers} encountered when employing ChatGPT to assist the programming experience. Third, we conducted a \textbf{comparative analysis} of how individuals use ChatGPT in comparison to SO, delineating their shared and distinctive affordances. Lastly, we discussed the design implications and offered a workflow for further development.

\section{Background \& Related Works}
This section offers an overview of past research on Q\&A platforms and GenAI for programming. \Space{practices, delving into their history and evolution. We then focused on research that is specifically centered around SO. Finally, we explored a relatively new research area related to Gen AI for coding.}


\subsection{Q\&A Platform for Programming Practices}
In the past few decades, social media has dramatically changed the landscape of software engineering, with Q\&A platforms as one of the most recent tools emerging around 2010's and widely adapted by programmers~\cite{storey2014r}. Having improved response time significantly compared to traditional communication methods, social media is attractive for software developers~\cite{mamykina2011}. The first prototype of Q\&A sites was established by Ackerman's Answer Garden~\cite{10.1145/91478.91485}, and gradually evolved and exposed to the general population. In 2005, both Reddit and Yahoo! Answers, 2 popular Q\&A platforms were released to the public, with Reddit still being one of the most popular social media platform as of 2023.
Software engineers quickly started to make use of these platforms, and established Stack Exchange Network which then transformed to Stack Overflow, a Q\&A website that focused on programming-related questions. Research has followed to learn more about this transition and revolution. Storey et al. well described the revolution of social media in software engineering from non-digital, digital to socially enabled (including Q\&A platform)~\cite{storey2014r}. Squire et al. conducted the first study to assess the quality of developer support provided by Q\&A sites such as SO, in comparison with previous tools used for developer support, such as mailing list~\cite{squire2015should}. According to Mamykina et al., some software developers started to believe that  Q\&A platforms such as SO had replaced web search/forums as their main source of finding answers to their programming problems~\cite{mamykina2011design}. With this trend happening, we are curious about how Q\&A platforms will continue to evolve to better fit programmers' needs. Hence we chose SO, one of the most popular Q\&A sites that concentrates on programming-related questions, as our proxy for this research.

\subsubsection{Stack Overflow} 
Created in 2008, SO quickly became the most widely used Q\&A platform for software engineers in the world with about 50 million visits monthly~\cite{nasehi2012makes,SEdata2023}.  SO is centered around nine design decisions: voting, tags, editing, badges, karma, pre-search, search engine optimization, user interface, and critical mass~\cite{treude2011programmers}. As such a well-developed and popular platform, SO triggered numerous researchers to conduct studies on this platform. Mamykina1 et al. examined what contributed to SO's success, specifically engaged with the founder of this platform~\cite{mamykina2011design}. As a question-initiated system, much research focuses on analyzing the questions asked on SO, with Treude et al.'s paper in 2011 being one of the most cited works around SO~\cite{treude2011programmers}. They summarized 7 categories of questions: how-to, discrepancy, environment, error, decision help, conceptual, and review. In addition to the high-level summary, researchers also summarized the detailed use cases of SO~\cite{insight2023,antelmi2023age},  discussed 'how to ask the right questions'~\cite{ahmad2018survey,yang2014asking}, as well as analyzing the characteristics of the answers like response time~\cite{mamykina2011design} and attributes of recognized answers~\cite{nasehi2012makes}.  Other aspects of SO like reward system\cite{wang2018users}, novice experience~\cite{chatterjee2020finding}, and social norms~\cite{cheriyan2021norm}, were also discussed in various studies. These studies shed light on different aspects of SO's functioning and user behavior. In addition to analyzing data and deriving insights, researchers have also developed tools to automate various aspects of this process, including classifier~\cite{beyer2018automatically}, datasets~\cite{baltes2018sotorrent}, and code search and recommendation tool~\cite{zagalsky2012example}. These tools contribute to the efficiency and comprehensiveness of SO research efforts.

\subsection{Generative AI for Coding}
Transitioning from human-powered Q\&A platforms, there is a growing trend of people increasingly embracing AI-powered tools to aid in their programming processes. According to SO's sentiment report in 2023, 70\% of the developers are already using or plan to use AI tools in their development process~\cite{sentiment2023}. This trend has been significantly influenced by the advancements in LLM, particularly publicly available tools. One noteworthy example is Github Copilot, which was announced on June 29, 2021. Copilot, usually considered as an LLM-based code generation tool~\cite{vaithilingam2022expectation}, offers various assistive features for programmers, including the conversion of code comments into executable code and auto-completion for code segments, repetitive sections, entire methods, and functions~\cite{copilot2023}.

Although it is a relatively new topic, researchers have already begun exploring programmers' experiences with LLM-powered code generation tools. For instance, Vaithilingam et al. conducted a within-subjects user study with 24 participants to understand how programmers use and perceive Copilot~\cite{vaithilingam2022expectation}.
While these code-generation tools offer helpful assistance, they cannot fulfill all the needs typically met by traditional question-based solutions. Alongside these tools, there has also been a rising adoption of \textbf{GenAI-powered chatbots} for coding, a method that closely parallels traditional Q\&A platforms. For instance, Rose et al. created the Programmer’s Assistant, a conversational prototype system, to explore conversational interactions rooted in code. 
Beyond small-scale research endeavors, the advent of ChatGPT spurred a larger trend of embracing AI chatbots for coding purposes. Launched on November 30, 2022, ChatGPT has made a substantial social impact. Users quickly began sharing their experiences and discussions on social media platforms, providing researchers with an opportunity to investigate user experiences. 

Given the novelty of this topic, researchers have just begun planning studies, but there were already many interesting works that we could learn from. One primary area of interest is testing the accuracy of ChatGPT-generated answers. Kabir et al. conducted a study using SO posts as input and assessed ChatGPT's accuracy in generating answers and found a 52\% inaccuracies~\cite{kabir2023answers}. Despite inaccuracies, researchers still found users' expressing willingness to use these GenAI tools~\cite{lau2023ban}.

Recognizing both the similarities and differences between human-powered and AI-powered Q\&A platforms, researchers have increasingly shown interest in comparing these tools and exploring the tensions between them. On December 5, 2022, just a week after the launch of ChatGPT, SO officially banned all ChatGPT-generated answers, stating that the unreliability of those answers would affect the trustworthy environment of SO~\cite{ban2023}. Borwankar et al. swiftly examined the repercussions of this restriction~\cite{borwankar2023unraveling}, and observed a decrease in programming-related questions and changes in expressions of the answers on SO following the restriction. Xue et al. expressed concerns about how LLMs might pose a threat to the survival of user-generated knowledge-sharing communities, potentially undermining sustainable learning and long-term improvements of LLMs~\cite{xue2023can}.  

Beyond these tensions between platforms, recent research has also begun to question ChatGPT's effectiveness in aiding coding tasks by comparing it with other methods. For example, Choudhuri et al. conducted an empirical study and discovered that ChatGPT did not enhance participants' productivity or self-efficacy compared to traditional non-GenAI online resources~\cite{choudhuri2024far}. Instead, it significantly heightened their frustration levels. 
While their findings highlight significant issues with GenAI coding assistance, these results are constrained by the small sample size of their lab-based experiment. We hold a more optimistic perspective and aim to observe long-term adoption of these tools in real-world settings to integrate the strengths of both methodologies. 
A similar view stemmed from Cheng et al., who discussed how online communities shape developers' trust in AI tools and explored ways to leverage community features to foster appropriate user trust in AI~\cite{cheng2022would}.
Rather than creating opposition, our aim is to explore opportunities for collaboration and improvement based on the unique affordances of each tool. To facilitate this discussion, we will compare ChatGPT with SO and propose strategies and solutions. We anticipate that analyzing Reddit posts will uncover long-term solutions adopted by programmers in real-world settings, providing valuable insights into how we might advance our approach.

\section{Method}
In this study, we scraped Reddit posts from multiple subreddits associated with ChatGPT, SO, and coding. After a data filtering process, we conducted a thematic analysis of the selected dataset. Combined with insights from previous literature, we subsequently present the Results section.

\subsection{Data Mining}
Using pushshift.io and PRAW API, we retrieved all Reddit posts from November 30, 2022 (the date of ChatGPT's release) to April 30, 2023, before breaking changes to the Reddit API happened in June 2023. Our analytical focus revolved around two principal themes: the use of ChatGPT in programming contexts, and a comparative exploration of ChatGPT and SO.

In our initial attempt, we chose queries such as \emph{'coding'} and \emph{'programming'} to scour all Reddit posts. However, this approach proved imprecise, as pinpointing programming-specific content through queries proved challenging. Consequently, we shifted our strategy from a site-wide search to a selection of several targeted subreddits—a method previously employed in related literature~\cite{gauthier2022will}.

For the first research question, we extracted all posts from the \emph{r/ChatGPTCoding}. Furthermore, we selected 11 prominent coding-related subreddits, including \emph{r/learnprogramming, r/AskProgramming, r/coding, r/programming, r/codinghelp, r/ compsci, r/cscareerquestions, r/ProgrammerHumor, r/openso-}\emph{-urce, r/ComputerScience}, and \emph{r/tinycode}~\cite{reddit2023}. To expand our dataset, we also incorporated content from the \emph{r/ChatGPT} subreddit by employing queries like \emph{'coding'} or \emph{'programming'}.

For our second research question, which centered on identifying posts that draw comparisons between ChatGPT and SO, we employed queries containing both \emph{'ChatGPT'} and \emph{'Stack Overflow'}. These queries were required to be present either in the selftext or the title. We intentionally did not confine our search to specific subreddits, aiming to cast a wider net for relevant content.

\begin{figure}[h]
  \centering
  \includegraphics[width=\linewidth]{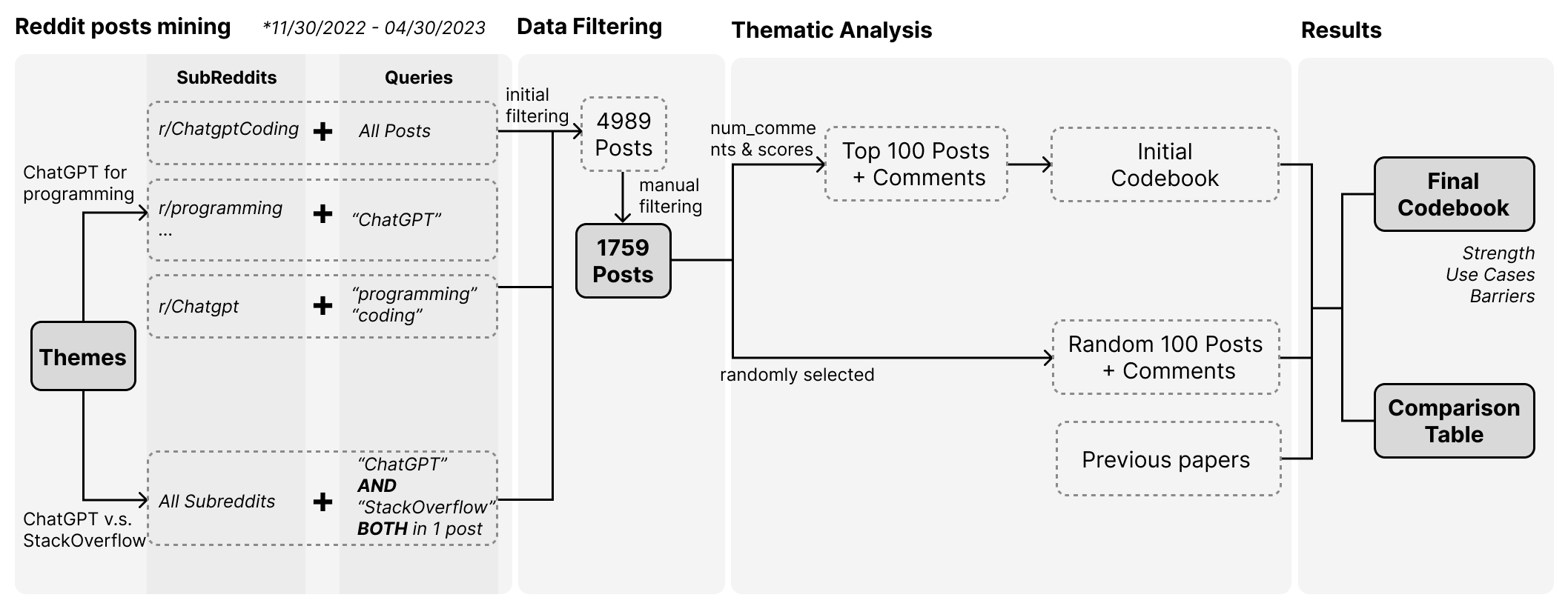}
  \caption{The workflow of data mining, filtering, and analysis.}
  \label{fig:method}

\end{figure}
 
\subsection{Data Filtering}
During the data exploration phase, we observed an overlap in the posts on the two identified themes. For example, many posts in \emph{r/ChatgptCoding} also discussed SO. Recognizing this pattern, we combined all the post data from different channels together as our final dataset. We then proceeded with initial data filtering such as eliminating duplicate posts. Subsequently, to enhance the precision of our dataset, manual filtering of irrelevant content was undertaken. This process involved researchers reviewing post content (selftext) and titles to ensure alignment. Ultimately, we reduced our dataset from over 4000 posts to 1,758 posts. The datasets are publicly available \footnote{\url{https://doi.org/10.6084/m9.figshare.24128880.v1}}. 

\subsection{Data Analysis}	
After obtaining the dataset, we employed an open coding approach on two research questions: ChatGPT for coding, and the comparison between ChatGPT and SO ~\cite{corbin1990grounded}. The codebooks were generated in two distinct phases: the initial development of a codebook and the subsequent refinement of the final codebook. Throughout the process, all authors discussed the codes and conducted codebook thematic analysis following the Grounded Theory~\cite{braun2006using, glaser1968discovery}.
Two researchers carried out the main coding process, discussing and refining the extracted codes iteratively until they reached a consensus to finalize the codebook. When contradictions arose, the coders engaged in joint discussions to refine the codebook by adjusting the definitions of codes or reinterpreting the data. All authors participated in the final discussion and refinement of the results presentation. Both coders have experience in programming and have light experience using Gen AI tools to assist in coding when they conducted the thematic analysis. Neither has posted on Reddit to discuss their experiences. Both coders have experience conducting thematic analysis before.
\subsubsection{Initial Codebook Development}\hfill\\
To construct the initial codebook, we selected the top 100 posts based on the highest number of comments and upvotes (scores). Given our aim to capture prevalent trends and recurring themes from the most discussed posts, the researchers examined not only the content of these posts but also delved into all the associated comments within each post~\cite{zhu2022empirical}.
\subsubsection{Final Codebook Development}\hfill\\
To enhance the comprehensiveness of our data analysis, we selected an additional 100 posts at random from the remained datasets. These selections, combined with the initial codebook, formed our final codebook. We achieved thematic saturation with our current sample size. By selecting both the top posts and random posts, we added an extra measure to avoid bias in popular posts. This approach is aligned with Ando et al.'s argument on bias from more communicative participants~\cite{ando2014achieving}. 

To address the first research question: \emph{In what ways are individuals employing GenAI
chatbots to enhance their coding experience?}, we conducted an inductive thematic analysis, {a bottom-up approach that allows themes and patterns to emerge naturally from the data, rather than being guided by predefined categories~\cite{braun2006using, glaser1968discovery}.
Following this method, we formed a codebook independently of prior research, then discussed the implications of our results and noted distinctions from previous studies in the Discussion~\cite{braun2006using}. To answer the second research question: \emph{what differentiates this coding experience from the utilization of conventional human-
powered Q\&A platforms?}, we conducted additional analysis on our final codebook. We reviewed previous works specifically related to the usage of Q\&A platform for coding assistance to drew the comparison table between those results and our codes. This comparison was also done after the coding process to avoid the temptation of applying existing theories to the coding process\cite{braun2006using}.
The key insights derived from this process are expounded upon in the subsequent section.
\subsection{Ethical Considerations}
All the Reddit posts collected and analyzed for the research were publicly accessible. We took care to remove any potential identifying information from the post content.
\section{Exploratory Quantitative Results}
Although our study primarily focuses on qualitative content analysis of Reddit posts, we begin by presenting an exploratory quantitative analysis to give our audience an overview of our dataset.
\subsection{Distribution of Posts}
The distribution of weekly post counts of the dataset, covering the period from November 30, 2022, to April 30, 2023, is illustrated in Figure \ref{fig:time}. Following a peak of discussions in the 1-2 weeks after the release of ChatGPT, indicating a consistently active data thread. The majority of these posts were shared within the \emph{r/ChatGPT} subreddit (38.9\%), followed by \emph{r/ChatGPTCoding} and several prominent programming subreddits [Fig. \ref{fig:time}].

\begin{figure}[h]
  \centering
  \includegraphics[width=\linewidth]{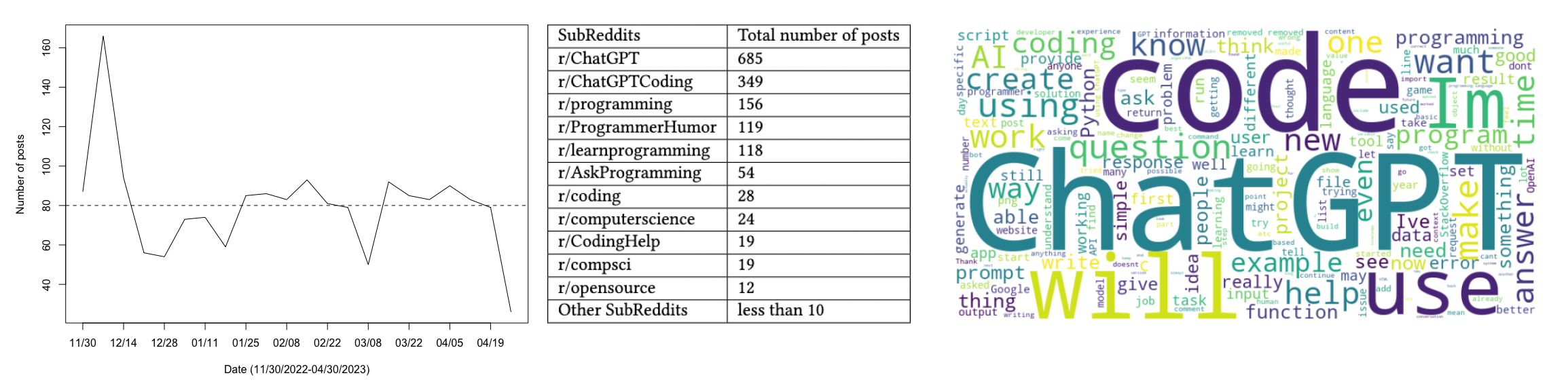}
  \caption{Description of the dataset: left: distribution of weekly post counts from 11/30/2022 to 4/30/2023; middle: distribution of post counts across various subreddits; right: wordcloud generated from the dataset}
  \label{fig:time}
\end{figure}

\subsection{Analysis}
First, we analyzed the posts through three primary measures, namely score,
num\_comment and upvote\_ratio, to obtain a basic descriptive statistic. The score represents the number of upvotes received by a submission, num\_comment indicates the number of comments on a submission, and upvote\_ratio refers to the proportion of upvotes a post or comment has received compared to the total number of votes it has received (both upvotes and downvotes). Our findings revealed that both score and num\_comment were positively skewed, with skewness values of 25.93 and 8.44, and upvote\_ratio was negatively skewed, with skewness value of -4.89. The medians were 55.21 (SD=689.01) for score, 11.67 (SD=42.99) for num\_comment, and 0.97 (SD=0.12) for upvote\_ratio. We also noted that a considerable proportion of the posts contain no comment (23.48\%), score (22.17\%), or both (5.00\%). These findings further validate our data analysis approach, which involved giving precedence to the most popular posts and supplementing them with randomly selected entries.

We generated a wordcloud\footnote{\url{https://github.com/amueller/word_cloud}} to visually represent the narratives across all the posts [Fig. \ref{fig:time}]. We also performed sentiment analysis on the posts using one of the state-of-the-art unsupervised sentiment prediction tools VADER\footnote{\url{https://github.com/cjhutto/vaderSentiment5}}~\cite{hutto2014vader}. The sentiment analysis of the posts reveals a moderately positive tone throughout the dataset, with an average compound score of 0.22. Among these posts, 37.91\% are categorized as positive, 53.70\% as neutral, and 10.13\% as negative. This distribution suggests that while the majority of the discussions are neutral or mildly positive, a significant portion still reflects varied sentiments, underlining the complexity of user experiences.}

\section{Qualitative Results}

In order to answer \emph{RQ1: In what ways are individuals employing GenAI chatbots to enhance their coding experience?}, we would like to discuss three themes: \textbf{Strengths}, \textbf{Use Cases} and \textbf{Barriers}. The comprehensive codebooks are available in the tables presented in the Appendix.

\subsection{Strengths}
Many individuals have reported highly successful experiences using ChatGPT to enhance their coding endeavors. In terms of the overall user experience, ChatGPT's adoption of natural language interactions contributes to an exceptional barrier to entry. One user expressed this by stating, \emph{"You will literally just describe the tool you need and you will get it."} Notably, even individuals who had previously given up on their programming pursuits expressed optimism about ChatGPT, believing it would help them \emph{"see it through this time."} This \textbf{gentle learning curve} underscores ChatGPT's potential in learning scenarios, particularly for novice users.

In regard to the answers provided by ChatGPT, users commend the platform for its {\textbf{fast responses}. They find the ability to receive a \emph{"quick and clear answer"} immediately after posting a question to be a highly satisfying experience. Compared to posting questions on SO, ChatGPT's responsiveness allows users to \textbf{maintain a state of flow} and bypass the need to overthink their inquiries, as the effort required to ask a question and obtain an answer is minimal. Users also appreciate the \textbf{clarity and detail} of ChatGPT's responses, with one user noting, \emph{"It even explains what each function did."} Moreover, interaction with ChatGPT reveals that it is not limited to providing a single question-answer interaction. Users can continue to engage ChatGPT to \textbf{refine its previous responses} or seek follow-up clarifications. This dynamic interaction comes into play when the initial response does not fully meet their needs (e.g. code contains bugs, system misinterprets questions) or when additional requirements arise within the same context (e.g. improving efficiency, adding new features). In instances where users still have difficulty comprehending solutions, they can directly request follow-up clarifications from ChatGPT.

Furthermore, software engineers who have explored ChatGPT have discovered its remarkable \textbf{customization capabilities} in generating responses. Some users have even asked ChatGPT to \emph{"explain certain code examples as if it were explaining to a 10-year-old"} to obtain more detailed and user-friendly answers. The user profiling capability of LLMs empowers software engineers to tailor answers to suit their specific needs. One user expressed their newfound confidence, stating, \emph{"Not sure if it's just because it is being explained to me in a specific way that I ask for, but I feel like I can better tackle my own coding projects after one day of use."}

In terms of the interpersonal aspect of interacting with ChatGPT, software engineers widely agree that ChatGPT fosters a more \textbf{respectful environment}, particularly when compared to SO. Users appreciate the absence of pedantic or condescending attitudes, with one user recommending ChatGPT to anyone \emph{"even slightly interested in coding."} People perceive ChatGPT as patient in providing answers, allowing them to ask repeated questions without concerns about receiving a response. Moreover, some liken ChatGPT to \emph{"an infinitely patient college professor who will politely and endlessly answer any and all questions you have,"} emphasizing the learning potential highlighted earlier in this section. A detailed table including the codes and quotes can be found in the appendix.

\subsection{Use Cases}

Now that we have established that many software engineers value interacting with ChatGPT to enhance their coding experiences, we will delve deeper into the specific use cases.

First, programmers frequently turn to ChatGPT for \textbf{direct assistance} when coding. Many individuals mention seeking ChatGPT's guidance on syntax, libraries, or specific functions. Another prominent use case involves users receiving code snippets or even entire scripts directly from ChatGPT. Users have also inputted paragraphs of code, asking ChatGPT to assist in \textbf{debugging}. Remarkably, some users have discovered that ChatGPT \emph{"comments the code fairly well,"} prompting them to employ it not just for writing code but also for \textbf{documentation} purposes. Following the completion of coding-related tasks, software engineers also use ChatGPT to aid in \textbf{testing} process. They utilize ChatGPT's 'generative' functionality to produce unit tests and manage edge cases.

Beyond assistance directly tied to specific coding tasks, individuals have explored ChatGPT's utility in generating \textbf{higher-level coding solutions}. Software engineers frequently employ ChatGPT to serve as a starting point helper, provide general direction, or suggest approaches to solving coding challenges. One user described this approach as the 'correct' usage of ChatGPT for coding: \emph{"It’s supposed to point you in a general direction and then you use something it doesn’t have: your brain."} Users also leverage ChatGPT to enhance code efficiency and even to facilitate the transfer of programs from one programming language to another.
An analysis of these use cases reveals potential applications for \textbf{learning} as well. When receiving instructions from ChatGPT for coding-related tasks, users occasionally encounter confusion, as highlighted in the earlier section. In such instances, users turn to ChatGPT to generate examples and provide explanations. \emph{"With chatGPT I can have it show me example code, explain it, and answer any questions with more examples."} For some, ChatGPT serves as a virtual pair programmer, with users asserting that both Co-Pilot and ChatGPT \emph{"won't replace us but augment our daily work."} A prominent post on r/learnprogramming explicitly underscores ChatGPT's capacity as a training tool: \emph{"Ask it questions like: 'Can you give me a set of recursive problem exercises that I can try and solve on my own?' And it will reply with a couple of questions, along with explanations if you're lost. Super neat!"} as stated by the author of the post.

Furthermore, users have explored intriguing applications of ChatGPT that broaden our perspective. Discussions have arisen about configuring ChatGPT to function as a code editor or other types of \textbf{simulators} nested within ChatGPT due to its \emph{"unbelievable programmability."} We anticipate that over time, there will be a proliferation of innovative use cases.

\subsection{Barriers}
While many individuals have reported successful experiences using ChatGPT to enhance their coding practices, there are also significant barriers and concerns that need to be addressed.

One of the most prominent issues with ChatGPT's answers is related to \textbf{reliability}. Specifically in coding scenarios, there have been instances where ChatGPT generated different fictitious codes each time it was asked the exact same question, undermining users' trust in the system. To the surprise of many programmers, ChatGPT often fabricated non-existent libraries, commands, or citations. These red herrings, or false leads, can be particularly detrimental as they place a significant burden on users to verify each piece of information, potentially eroding trust in the entire solution. Users have also noted that even when ChatGPT provides incorrect or ambiguous answers, it does so with a \emph{"calm and assured confidence,"} highlighting a disparity between ChatGPT's responses and users' expectations, especially when users begin to treat it as a real assistant and apply human-like judgment. This unreliability is exacerbated by the fact that ChatGPT operates as a standalone program. As one user put it, \emph{"ChatGPT only helps if you can judge whether its answer is correct, unlike an answer with 50 upvotes on SO, which you \_know\_ is correct."}

During their exploration, users discovered that ChatGPT's responses were \textbf{sensitive} to the way questions were phrased, with better answers often requiring a clear and specific goal. While prompt engineering has long been recognized as a crucial aspect of developing LLM, software engineers unfamiliar with this process found it time-consuming to learn how to write effective prompts: \emph{"Anyone can type something in and get an answer, but getting the answer you want is a different story."}

Another significant challenge faced by ChatGPT is the \textbf{lack of transparency}. Frequently, there are no citations or links to supporting documentation, and in some cases, ChatGPT even generates fake documentation. This lack of transparency adds an additional layer of difficulty in validating the accuracy of answers. Furthermore, ChatGPT is often ambiguous about its training resources, and there have been instances where it provided false information about its training data.

When users attempted to use code generated by ChatGPT, some realized potential issues related to \textbf{code management}. Some expressed concerns that ChatGPT \emph{"will likely create code that is not maintainable,"} while others worried that it disrupted coding styles, with one user stating, \emph{"It looks like someone coded it with a split personality."}

Unsurprisingly, users have encountered limitations in terms of \textbf{resource access}, including constrained training sets, the absence of live resources after 2021, the inability to run ChatGPT locally, and the lack of integrated development environment (IDE) plugins. Additionally, there have been instances where ChatGPT refused to generate code, especially when specific trigger prompts were used. Users have also expressed the need for an 'adult' version of ChatGPT with fewer restrictions, as many of the current limitations are related to ethical concerns.
Other access constraints include rate limitations, regional restrictions, financial implications, downtime, and system overload. Additionally, some users expressed concerns about \textbf{copyright} issues related to using ChatGPT's generated code, particularly in professional settings.

Having gained a more comprehensive understanding of how ChatGPT is employed to enhance the coding experience, we proceed to our second RQ: \emph{\textbf{R2: What differentiates this coding experience from the utilization of conventional human-powered Q\&A platforms?}}.
\subsection{Comparison: SO vs. ChatGPT}
Building upon the themes outlined in previous sections and informed by a thorough literature review, we conducted a comparative analysis of these two tools across various dimensions.

ChatGPT exhibits a notably \textbf{faster response time}, typically within a few seconds, whereas the average response time for a SO post is approximately 11 minutes~\cite{mamykina2011design}. This disparity holds significant implications for their use cases described previously. One user highlighted this, stating, \emph{"I think it's a better resource than Stack, it answers questions fast, \textit{keeping you in a flow state} all with no attitude about how to ask a question."} This observation was also tied to the acceptance of \textbf{repeated questions} on SO. SO's code of conduct includes guidelines on how to ask a question~\cite{conduct2023}, and it discourages posting duplicate questions, requiring users to engage in additional searching before posting a question. In contrast, ChatGPT allows users to ask the same question multiple times, eliminating the need for extensive pre-searching and enabling direct access to solutions at any time. This affordance can be particularly advantageous for seeking information related to relatively common and straightforward queries, such as those pertaining to syntax and libraries. Users have even recognized the potential for ChatGPT to replace certain features provided by search engines. One user commented, \emph{"If anything, ChatGPT has made me realize just how inefficient Google is. Because clearly, this information is out there on the net somewhere. But Google sure as hell can't retrieve it. Google instead gives me the top link as a thread in which the only answer is 'this is a duplicate of some other thread,' with a link to that thread. And that was the only tool we've had to find this kind of information for decades."} Similarly, users have expressed that ChatGPT's low learning curve in formulating questions has facilitated their interactions, with one noting, \emph{"I've asked it questions when I can't think of how I would word my Google search and have received good results."}

The \textbf{iterative approach} plays a vital role in the Q\&A process and is a cornerstone of successful problem-solving practices~\cite{d1971problem}. In the context of Q\&A processes, both SO and ChatGPT support iteration, albeit in different ways - SO employs comments within the same post~\cite{zhu2022empirical}, while ChatGPT facilitates follow-up questions within the same conversation. However, a potential distinction arises in that many SO posts tend to be problem-focused. SO's Code of Conduct(CoC)~\cite{conduct2023} allows users to add comments for purposes such as requesting clarification from the author, offering constructive criticism, or appending relevant but \textbf{minor or transient} information to a post~\cite{conduct2023}. It discourages users from posting comments to initiate secondary discussions. While this policy maintains the cleanliness and organization of SO as a public platform, it may limit its ability to provide more personalized assistance to individual users.

In terms of providing \textbf{detailed explanations}, there is a similarity between SO and ChatGPT. ChatGPT typically offers detailed explanations, and on SO, this factor is instrumental in posts gaining popularity~\cite{treude2011programmers}. Concerning the level of \textbf{customization}, both platforms support a basic level of customization, enabling users to add context to their questions. Additionally, users on both SO and ChatGPT exhibit a diligent effort to carefully read questions. However, ChatGPT has a distinct advantage in user profiling capabilities, allowing it to customize the way it delivers explanations to users. SO, being a social community, strives to maintain consistency in its replies and may have less flexibility in adjusting narratives to cater to each requester's specific needs. As an automation system, ChatGPT has greater potential for extensive customization.

The recurring reference to SO's CoC~\cite{conduct2023} underscores the importance of adhering to \textbf{social norms} when posting questions on SO. Failure to follow the CoC can result in users' questions going unanswered or even being deleted~\cite{conduct2023}. In contrast, while ChatGPT does not close users' questions, obtaining desired answers may still require some effort, as we will elaborate later.

As mentioned previously, the issue of \emph{"how to ask a question"} is crucial for both platforms, affecting not only adherence to norms but also the efficiency of receiving correct answers. Previous research suggests that concrete, specific, and clear questions tend to elicit better answers~\cite{yang2014asking, ahmad2018survey}. Similarly, users have reported that providing a clear and specific goal assists ChatGPT in generating responses. However, further research is needed to validate this assertion and to compare the differing \textbf{input requirements} of the two platforms for achieving better answers.

Even when users pose well-constructed questions, there are still questions of \textbf{reliability}. The reliability of ChatGPT vs. SO has been a topic of debate. In Background, a study by Kabir et al. input questions from SO to ChatGPT and found that 52\% of ChatGPT’s answers contain inaccuracies, while users still prefer ChatGPT’s responses 39.34\% of the time~\cite{kabir2023answers}. Opinions on ChatGPT's accuracy varied in our dataset. Some users described ChatGPT as \emph{"clear, confident but wrong,"} while others asserted that \emph{"ChatGPT answers curated by humans are already better than the average SO answer, in my humble opinion (IMHO)."} A more comprehensive analysis is necessary to compare the accuracy of the two tools, particularly given the different ways people ask questions on these two platforms. Nevertheless, one significant advantage of SO is its upvote and selection features, providing an additional resource for validating the reliability of answers through crowdsourcing. 

Regarding \textbf{resource access and the features}, SO, as a crowd-powered platform, can collect updated information but is limited by its user population, which has not been problematic thus far. ChatGPT, despite being an LLM trained on massive datasets similar to crowdsourced resources, is often confined to outdated training sets, a limitation that could be addressed in the future.

We noted that many Reddit posts were based on personal experiences, but one consensus emerged: ChatGPT was perceived as significantly more \textbf{user-friendly} than SO. SO has a long reputation for being toxic. \emph{"The fact is, SO has a serious problem with bullying, one that SO itself has been trying really hard to mitigate, without success so far."}, from one Reddit post. In contrast, when using Chatgpt, users did not need to worry about asking \emph{"stupid"} or repeated questions. 

For \textbf{transparency}, SO, as an open-source platform, upholds a high degree of transparency. Answers that include links to supporting documentation have been highly appreciated by users~\cite{conduct2023}. In contrast, ChatGPT exhibits notable weaknesses in terms of transparency, both with its training datasets and the answers it generates. In terms of \textbf{accessibility}, beyond common issues such as network access, ChatGPT faces more challenges than SO, including rate and regional restrictions.

In light of these differences in asking questions and providing answers, were there any significant variations in the practical \textbf{use cases} observed by users? Both platforms, operating as question-based platforms, shared many similarities in use cases. The use cases and initiatives described in the works of Treude and Nasehi~\cite{treude2011programmers,nasehi2012makes} align closely with the scenarios discussed in the Reddit posts. However, it's notable that ChatGPT serves more as an assistant within the user's workflow, particularly in learning contexts, such as functioning as a pair programming partner or generating quizzes. Further investigation is warranted to explore the potential variations in their use cases, especially if access to ChatGPT's log datasets becomes available.

In \textbf{summary}, compared to SO, ChatGPT exhibits a faster response rate, a gentle learning curve, and higher tolerance, qualities that are particularly advantageous for addressing \emph{"trivial"} or common questions. Both platforms support iteration, with ChatGPT being more tolerant of topic switching within a single conversation. Both SO and ChatGPT provide detailed and customized explanations, although ChatGPT has greater potential for extensive personalization such as user profiling. SO enforces more norms and a CoC compared to ChatGPT. Further studies are needed to investigate the different input requirements and reliability of the two platforms, with SO benefiting from additional validation techniques such as voting. ChatGPT is perceived as significantly more user-friendly than SO. SO excels in transparency and accessibility. Concerning specific use cases, despite the large overlaps, ChatGPT demonstrates its extensive potential as an assistant, particularly in learning scenarios. A visual comparison of the two platforms is also presented in Table. \ref{tab:compare}.

\begingroup
\renewcommand\arraystretch{1.3}
\begin{scriptsize}
\begin{longtable}{lll}
\toprule\toprule
\centering
Themes                     & ChatGPT                    & SO                    \\ 
\rowcolor{Gray}
\textbf{Response Speed}             
& 
Several seconds            
& 
Average 11 minutes\text{~\cite{mamykina2011design}}    
\\

\rowcolor{Gray}
\textbf{Accept Repeated Questions?} 
& 
Yes                        
& 
Low tolerance         
\\

\textbf{Iterative Approach} 
&
Follow-up questions 
&
Discussion in comments \text{~\cite{zhu2022empirical}} 
\\
\textbf{Detailed Explanations}  &
  Yes 
  &
  Popular answers \text{~\cite{treude2011programmers}} 
  \\
\textbf{Extensive Customization}    
& 
Enable user profiling      
& Yes                   
\\
\rowcolor{Gray}
\textbf{Norms} &
  Natural language &
  Code of Conduct \text{~\cite{conduct2023}}
  \\

\textbf{Input Requirement} &
  Clear goal, and? &
  Concrete \text{~\cite{treude2011programmers}}, specificity, and clarity \text{~\cite{yang2014asking}} 
\\
\rowcolor{Gray}
\textbf{Reliability}                
& 
Clear, confident and wrong 
& 
Additional validation 
\\

\textbf{Resource and Feature}       
& 
LLM                        
& 
Crowd power 
\\
\rowcolor{Gray}
\textbf{Respectful Environment}
&
Yes 
&
'Bullying' culture \text{~\cite{notwelcoming2018}} 
\\
\rowcolor{Gray}
\textbf{Transparency \& Accessibility}
&
Weak 
&
Good
\\
\textbf{Use Cases}
&
Personal assistant + Learning potential 
&
7 categories\text{~\cite{treude2011programmers}}
\\
\bottomrule\bottomrule
\label{tab:compare}\\
\caption{Compare the distinct characteristics of SO and ChatGPT in supporting coding practices. Notable differences are indicated in gray.}
\end{longtable}
\end{scriptsize}
\endgroup
When comparing the two platforms, we noticed similarities and differences across various aspects. In some cases, they even complement each other. How do these findings differ from previous work, and how can we leverage the strengths of both to design effective solutions? We will explore these questions in more detail in the following section.

\section{Discussion}
In this paper, we explored the usage of ChatGPT in assisting coding practices, particularly compared to SO. We focus on understanding the strengths, use cases, and barriers of ChatGPT and conducting a comparative analysis with SO. In this section, we compare our findings with prior work. \Space{We present design implications, including general principles and specific solutions to combine the strength of Gen AI-powered chatbot and human-powered Q\&A platform. 
After that, we delved into our perspectives on the transformative impact of Gen AI on coding assistants, and described the limitations of our study.}
\subsection{Long-Term Adoption of GenAI in Coding: Reddit Posts Versus Lab Experiments}

Our results complement experimental studies of coding experiences with ChatGPT and Copilot~\cite{choudhuri2024far,vaithilingam2022expectation}.
\Space{In the previous section, we discussed two highly relevant studies that examined coding experiences with the assistance of ChatGPT and Copilot in experimental settings~\cite{choudhuri2024far,vaithilingam2022expectation}. While our focus is on long-term adoption in real-world settings, as evidenced by Reddit posts, how do the results compare and contrast with those from the studies?}
\Space{We observed several similar patterns.}
Prior work has also noted that Gen AI assistants excel at providing a starting point, are easy for beginners to use, and save programmers' time. 
\Space{Both studies highlighted issues such as reliability and errors, poor code maintenance, and sensitivity to input with these tools.}
Perhaps the most interesting findings, however, are around differing results and conclusions.
Choudhuri et al. reported negative outcomes from using ChatGPT for coding assistance, noting that it neither enhanced productivity nor self-efficacy and instead increased frustration~\cite{choudhuri2024far}. 
However, in our analysis of Reddit posts, sentiment analysis revealed moderate positivity and the qualitative feedback from users presented a more mixed picture with both strengths and barriers.

Long-term users might have had more time to adapt and find effective ways to utilize the tool, developing specific prompting strategies like \emph{``explain certain code examples as if it were explaining to a 10 year old''} mentioned in one of the posts. 
These prior studies focused on task completion, which, while important, does not capture the long-term sentiments about these tools, which we were able to identify in the posts. 
In our findings, individuals who have used ChatGPT for coding assistance in real-world scenarios greatly appreciated the respectful and patient environment it fostered, contrasting with their experiences on platforms like SO. Access and copyright issues, often overlooked in lab studies, are incredibly crucial and significantly impact long-term usage, especially in professional settings.

Programmers have more time to explore different uses of the tools in the wild, not restricted to task completion as dictated by specific study designs. 
Many Reddit posts indicate that programmers use ChatGPT for coding-adjacent activities, such as generating documentation, unit tests, and managing edge cases. 
Other use cases, such as learning, could only be captured in long-term usage with ChatGPT, where users viewed it as a coding assistant and pair programming partner, and even used it to generate quizzes. 
Similarly, other emergent use cases, such as employing it as a code editor or for simulating operational systems, can be detected using our methodology.

While the analysis of Reddit posts reveals additional insights not found in laboratory settings, we also recognize the limitations of this method. 
Posts submitted by programmers are usually asynchronous, which misses capturing their immediate reactions and impedes our opportunity to conduct deeper quantitative analysis. For instance, Vaithilingam et al. captured the cognitive load in using Copilot, something programmers might not recall and discuss in their posts. Additionally, our approach can not measure productivity metrics or conduct comparative analyses like Choudhuri et al. As a result, we believe our work effectively complements existing research on empirical experiments. Recognizing the differences in results, we encourage researchers to explore both short-term and long-term usage of GenAI tools for coding to capture comprehensive user feedback.

\subsection{Future Design Implications}
Recognizing the commonalities and differences between these two platforms, our objective is to enhance the design GenAI coding assistants to better address users' needs by leveraging the strengths of both systems. Drawing from the insights gained through previous findings, we derive several design implications for future GenAI-powered chatbots designed to assist with coding tasks.
\subsubsection{General Implications and Actions}
To maintain the current strengths of ChatGPT, future AI-powered chatbots for coding assistance should follow these general implications and actions:

\noindent\textbf{Implication: Comprehensive}

\noindent\emph{Action: Provide detailed explanations with examples of code snippets}

\noindent\textbf{Implication: Customized}

\noindent\emph{Action: Tailored to the coding languages, problem-solving context, and user expertise in coding}

\noindent\textbf{Implication: Patient}

\noindent\emph{Action: Consistently provide answers when coding questions are simple and repetitive (e.g. syntax)}

\noindent\textbf{Implication: Reliable, Transparent and Explainable}

\noindent\emph{Action: Offer supporting documentation (e.g. libraries, citations, links) and ensure they are accurate}

\noindent\emph{Action: Self-identify and report its reliability level to the user}

\noindent\emph{Action: Introduce external validation check outside of a single standalone platform}

\noindent\textbf{Implication: Consistent}

\noindent\emph{Action: Always built on previous answers.}

\noindent\textbf{Implication: Aligned Tone}

\noindent\emph{Action: Avoid overconfidence when reliability is low}

We recognize that our list of design implications aligns with previous works, particularly in terms of explainability within Explainable AI (XAI) and some discussions around the consistency of LLMs~\cite{beaulieu2024evaluating,zhou2023red,arrieta2020explainable}. While detailed suggestions for action require further verification through additional studies, our Reddit analysis provides additional insights into potential solutions. 
We find that users are more concerned with whether the results from GenAI effectively assist with their coding tasks rather than the underlying mechanisms of the AI system.
As a result, we suggest providing explanations and validations focused on the results rather than on the model itself~\cite{dong2022towards,ghaeini2019saliency}. Hallucination and inconsistency are often identified as fundamental issues with LLMs that are difficult to resolve. While discussing technical solutions for LLM issues is beyond the scope of this study, we believe that altering the presentation of the results may mitigate these problems. 
Instead of focusing solely on improving the model's explainability, providing evidence such as libraries, citations, and links along with the results is crucial for users to assess their reliability.
Another useful solution proposed by users is to have GenAI report the reliability and confidence levels of its results. 
Introducing external human validation is another solution, which we will discuss more thoroughly in the next section. 
Inconsistency is a fundamental issue with LLMs that is difficult to resolve, and we noticed it becomes particularly problematic for users when they ask follow-up questions about the same issue. Consequently, building on previous answers and using them as prompts for subsequent queries, rather than generating entirely new solutions, is a temporary measure that may enhance user experience.

\subsubsection{Different Strategies According to Scenarios}
We recognized that real-world practice presents different scenarios with distinct design considerations. In this section, we will delve deeper into the detailed design strategies of how the LLM system should respond to various questions.

\noindent\textbf{Discern the scenarios}\indent Before formulating strategies, the system should first be capable of discerning the coding scenario. This can be achieved through various means: \textbf{understanding contextual information in users' questions, proactively inquiring about their objectives, or even utilizing external materials} such as linking with a code editor to gather more information about the coding purpose. Based on our analysis, there are primarily three distinct scenarios in which we will discuss the design implications for each.

\noindent\textbf{Scenario 1: Providing an answer}\indent One of the most common scenarios for using ChatGPT in coding is to enhance the completion of a coding task. In such cases, users aim to acquire sufficient information to complete the task, typically seeking a 'correct' answer for each question. Consequently, the primary objective of the LLM system is to ascertain that answer and present it in a manner that allows users to swiftly proceed with their task. A frequent task in this context involves clear instructional inquiries, such as those about syntax/libraries/functions. In such scenarios, the key goal is to provide a quick and clear answer, enabling users to promptly return to their original task. Detailed examples might not always be necessary, depending on the users' expertise. This task often resembles replacing the traditional function of a search engine.

\noindent\textbf{Scenario 2: Problem-solving}\indent However, in other situations, the task quickly evolves from obtaining a simple answer to engaging in complex problem-solving. Debugging serves as a prime example; it often requires multiple iterations to pinpoint errors in the code. In such scenarios, GenAI needs to collaborate with programmers to resolve a problem, involving a more natural conversational back-and-forth process. 
Outlining a coding project also fits into this category. In these cases, providing detailed explanations and external materials as validity verification becomes necessary to build trust and reach consensus. We recognize that such processes are similar to those employed in online Q\&A platforms (e.g. comments, ratings, selecting correct answers, etc.). A future GenAI system could learn and emulate many strategies from platforms like SO to foster collaboration and problem-solving, rather than solely focusing on generating correct answers.

\noindent\textbf{Scenario 3: Learning}\indent Lastly, learning is another important scenario for using GenAI as a code assistant. In such cases, providing highly detailed explanations and examples with documentation becomes crucial, not only for transparency but also to offer users more materials in their learning journey. The tone of the answers also becomes more important; it needs to be patient and supportive, forming a respectful learning environment. Extensive customization becomes crucial since it involves more than just addressing the context of the problem but supporting personal growth. GenAI system should be able to adapt its content extensively based on the users' growing expertise.

\subsubsection{Combine Human Power with Gen AI}

In our effort to compare SO with ChatGPT, we identified several specific drawbacks associated with SO. These include slower response times, a lower tolerance for questions deemed less impactful (e.g., repeated, trivial, too personalized), or those that could potentially hinder the platform's cleanliness as a social media platform (e.g., independent follow-up questions), and the presence of an unfriendly culture. It became evident that many of these challenges could potentially be mitigated by harnessing the capabilities of AI, which may contribute to the ongoing shift in coding assistance practices. Rather than presenting future design improvements for SO, we propose a workflow that leverages the strengths of both platforms.

SO, functioning as a blend of Q\&A and social media, differs in its operation from ChatGPT. We acknowledge the potential disparities in usage phases between these two platforms. ChatGPT, being an `always-on' personal assistant, serves as the initial phase of assistance. Developers could get fast and clear answers without too much effort. Upon receiving answers from AI after several iterations, in cases where the problem remains unsolved or users harbor doubts about response accuracy, we propose introducing external validation and assistance. In Phase II, developers seek crowd guidance by directly reporting issues to Q\&A platforms like SO. Other software engineers can then participate in the decision-making process by voting to verify answer accuracy, providing comments, and offering additional answers to both the initial and follow-up questions. Given that SO banned AI-generated answers partly due to identification difficulties, we suggest introducing specific tags like "AI-generated", for posts to enhance clarity and categorization.

After obtaining answers from the Q\&A platform, we can continue to harness the extensive customization capabilities of the AI chatbot as a personal assistant to further tailor responses. In Phase III, this approach lets AI handle the detailed personalization process, maximizing the strengths of both AI and crowd communities while maintaining the social platform restrictions.

In this workflow, AI chatbot handles tedious or too personalized tasks, while Q\&A serves as an external assistant that maintains its own affordances as a social platform. While this workflow is by no means perfect, we aspire to offer inspiration on how future systems could potentially harness the strengths of both platforms to further enhance developer assistance in the coding process.

\begin{figure}[h]
  \centering
  \includegraphics[width=0.6\linewidth]{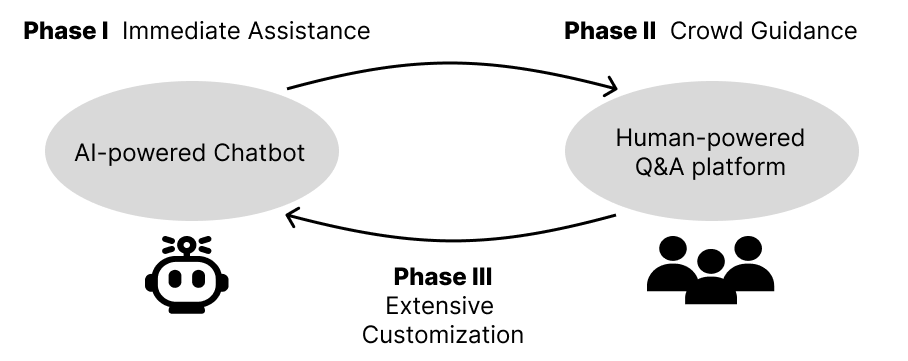}
  \caption{Different phases of seeking coding guidance through AI-powered chatbot and Q\&A platform.}
  \label{fig:phase}

\end{figure}

Combined with Sec 5.4 [Table. \ref{tab:compare}], we summarized potential solutions for each tool's issues using the combined workflow we described \emph{(QA as human-powered Q\&A platform)}:

\noindent\textbf{Problem: Lack of reliability and transparency in answers from GenAI}

\noindent\emph{Solution: Introduce external validation from QA to evaluate AI-generated content (e.g. voting).}

\noindent\textbf{Problem: Slow response and low tolerance for trivial or repetitive questions in QA}

\noindent\emph{Solution: Route simpler, more repetitive inquiries to GenAI for quicker and more patient responses.}

\noindent\textbf{Problem: Reluctance to post follow-up questions on QA platforms (CoC)}

\noindent\emph{Solution: Use GenAI as a personal assistant to answer follow-up questions and reduce the use of QA (e.g. comments) to keep it cleaner and focused on valuable questions that benefit the wider public.}

\noindent\textbf{Problem: Inability to customize responses from QA according to personal preferences}

\noindent\emph{Solution: Pass answers from QA through GenAI for further customization and detailed explanations.}

\noindent\textbf{Problem:  Challenges in accessing Gen AI tools.}

\noindent\emph{Solution: Leverage traditional crowd-sourced assistance.}

\subsection{From the Past to the Future: The Evolving Role of GenAI in Coding Assistance}
In the Background section, we referenced Storey et al.'s depiction of the evolution of software engineering with social media, progressing from non-digital to digital and, ultimately, socially enabled platforms like Q\&A forums~\cite{storey2014r}. The use of ChatGPT signifies another transformation in coding assistance. Despite these new technological tools, the fundamental needs of programmers remain largely consistent—finding more efficient ways to solve problems. Questions posted on any platform may not always be detailed enough, and code snippets sourced from various platforms may not be flawless. As one post humorously noted: \emph{"Back in my time, there was no SO, only trial and error. Mostly errors, though. I guess some things never change."} Nonetheless, GenAI introduces unique opportunities for its generality. In the past, humans generated content and organized them for easy retrieval. Now, even the content generation aspect can be partially automated with AI. What implications might this hold for software engineers? Many initially consider potential threats. Numerous Reddit posts ponder whether ChatGPT could potentially replace human programmers, or if proficiency in using ChatGPT might become a crucial skill for software developers in the future. These debates are ongoing, and while the future remains uncertain, there are ethical considerations that need to be addressed to safeguard individuals' rights in future studies.

As highlighted in the previous section, we have recognized ChatGPT's significant potential in learning scenarios. Coding, being a distinct type of task, has witnessed its learning process become publicly accessible due to the rise of open-source culture, albeit often in a fragmented manner. This presents an excellent opportunity for GenAI to leverage extensive open-source materials and generate instructional content. Furthermore, chatbots' ability to adopt personas positions ChatGPT as a versatile personal assistant, capable of serving as a tutor, grader, pair programmer, and more. However, this potential has sparked debates within the community. One such debate centers on the concept of \emph{"true learning"} in coding. Some argue that it entails problem-solving, memorization, understanding code mechanisms, reading documentation, and acquiring lasting knowledge. Conversely, others contend that these goals must be contextualized in real-world scenarios, where copying and pasting code are integral aspects of the learning process. Exploring how ChatGPT can aid in achieving these learning objectives presents an intriguing area of study.

In the previous section, we presented a sample workflow for seeking programming guidance through both AI-powered chatbots and human-powered Q\&A platforms. We aimed to leverage the strengths of both platforms, but we also pondered whether, with researchers continually enhancing AI to harness human creativity and tackle complex tasks, it might eventually replace platforms like SO and other Q\&A sites. While it's impossible to predict the future with certainty, we do believe that AI will play a significant role in replacing some human support. However, it's crucial to note that current AI systems still learn from human-generated data, even for GenAI which can create content. As demonstrated by Xue et al., the lack of user sharing open-sourced information could pose challenges to the sustainability of LLM's learning process~\cite{xue2023can}. Therefore, we suggest that future systems strike a balance, not only for the sustainable improvement of AI but also to preserve the social aspects tied to Q\&A platforms. These platforms should continue to be spaces where users not only seek answers but also engage in a community to collectively find solutions.

\subsection{Limitations}
While we aimed to ensure the robustness of each research step, there are still some limitations to this work. Although ChatGPT has garnered considerable attention during the specified timeframe, comprehensively capturing individuals' complete experiences with it through Reddit post analysis remains challenging due to factors such as the silent effect and potential biases~\cite{dark2012}. Our data analysis strategy has been carefully devised to accurately capture the prevailing trends, but augmenting our resources could lend greater robustness to our results. In the future, conducting follow-up comparative user studies would be valuable to further validate our insights.

With enhanced resources, addressing the technical intricacies could involve refining the search process by extending it beyond the posts' titles and selftext to encompass the entire content, including comments. 
Moreover, the main changes in Reddit APIs and the shutdown of pushshift.io impacted our ability to access more expansive datasets. 
Readers should also be aware that the accuracy of the API is not guaranteed to be 100\% primarily due to potential delays~\cite{praw2023}. Nonetheless, this approach remains the most accurate within the scope of research~\cite{gauthier2022will}. It's worth noting that our results have exhibited a greater degree of promise since our emphasis on qualitative analysis.

As we narrow our focus to chatbots, we've acknowledged the presence of other GenAI tools like Co-pilot~\cite{copilot2023}. Our current research centers on Q\&A platforms so we chose ChatGPT as a chatbot for the comparison, but there's potential for future studies to encompass broader AI-powered assistance. Additionally, we chose SO and ChatGPT as representative examples of the two discussed categories. Although this choice is backed by existing literature, we're mindful of other tool's existence and have carefully confined our takeaways to prevent over-generalization of our findings.

\section{Conclusion}
In this study, we embarked on an exploration of the experiences of software engineers as they engage with generative AI chatbots versus human-powered Q\&A platforms in their coding endeavors. We employed Stack Overflow (SO) and ChatGPT as representative platforms, and through a rigorous thematic analysis of Reddit posts, we gained insights into the integration of these two forms of assistance into the real-world coding experiences of programmers, including strength, use cases and barriers. This exploration serves as a springboard for future investigations into the evolving landscape of AI-driven coding aids, as we strive to unlock their full potential in empowering software engineers on their coding journeys.
\section{Acknowledgment}
We are grateful to our collaborators for all their help and support throughout this project. Their knowledge and guidance shaped our research and improved its overall impact. We also appreciate the reviewers’ careful reading and thoughtful comments, which prompted us to clarify our ideas and present our findings more effectively.
\bibliographystyle{ACM-Reference-Format}
\bibliography{bio}
\received{January 2024}
\received[revised]{July 2024}
\received[accepted]{October 2024}
\newpage

\appendix
\section{Codebooks with Themes and Quote Examples}

\begin{scriptsize}
\renewcommand\arraystretch{1.2}
\begin{longtable}{lll}
\caption{Example themes, sub-themes and quote examples that demonstrate the \textbf{Strengths} of using ChatGPT to assist coding from Reddit posts.}
\label{tab:strength}\\
\toprule\toprule
Themes& Sub-Themes                & Quote Example \\
\midrule

\begin{tabular}[t]{@{}l@{}}\textbf{Gentle Learning} \\ \textbf{Curve}\end{tabular} &                           
&  
\begin{tabular}[t]{@{}l@{}l@{}}
\emph{"As someone who tried to learn programming once and gave up, having ChatGPT this second}\\
\emph{time around will hopefully help me see it through this time."}
\end{tabular}
\\
\midrule

\textbf{Fast Response}                &                           
&   
\begin{tabular}[t]{@{}l@{}l@{}}
\emph{"I think it's a better resource than Stack, it's answers questions fast, keeping you in flow state} \\\emph{all with no attitude about how to ask a question."}
\end{tabular}
\\
\midrule
\begin{tabular}[t]{@{}l@{}}\textbf{Detailed and Clear} \\ \textbf{Explanations}\end{tabular} 
&                           
& 
\begin{tabular}[t]{@{}l@{}}
\emph{"But slowly I'm making some progress thanks to the clear explanations provided. "}\\
\emph{"It even explains what each function did."}
\end{tabular}               
\\
\midrule
\textbf{Iterative Approach}                       & 
\begin{tabular}[t]{@{}l@{}}Refine previous\\ answers\end{tabular} 
& 
\begin{tabular}[t]{@{}l@{}}
\emph{"I have been copy and pasting my code into ChatGPT and asking it to make it more efficient,}\\
\emph{and after a few tries it comes up with some beautiful ideas."}
\end{tabular}
\\
&&\\
& 
\begin{tabular}[t]{@{}l@{}}Follow-up\\ clarification\end{tabular}
&   
\begin{tabular}[t]{@{}l@{}}
\emph{"If you don’t understand one of its responses, just respond with “can you elaborate further”, and}\\
\emph{it will tryexplaining it in more detail/in a different way."}
\end{tabular}
\\
\midrule
\begin{tabular}[t]{@{}l@{}}\textbf{Extensive} \\ \textbf{Customization}\end{tabular}                          &                           
&  
\begin{tabular}[t]{@{}l@{}}
\emph{"I'm a noob so I ask it to explain certain code examples as if it were explaining to a 10 year old,}\\
\emph{that helped a lot."}
\end{tabular}
\\
\midrule
\begin{tabular}[t]{@{}l@{}}\textbf{Respectful} \\ \textbf{Environment}\end{tabular} & 
No bullying               
&   
\begin{tabular}[t]{@{}l@{}l@{}}
\emph{"The fact that ChatGPT can almost completely replace Stack Overflow should encourage anyone} \\
\emph{who is even slightly interested in coding to give it a try, you won't have to deal with pedantic or}\\
\emph{condescending people."}

\end{tabular}
                          
\\
\midrule
\textbf{Increased Patience}                       & 
\begin{tabular}[t]{@{}l@{}}Accept repeated\\ questions\end{tabular}
&       
\begin{tabular}[t]{@{}l@{}l@{}l@{}}
\emph{"Compared to trying to post the same question with the skeleton code to Stack Overflow, the} \\
\emph{experience was like night and day. It would have been closed as a fake duplicate, or "needs more}\\ 
\emph{context", or some other bullshit reason a power tripping neckbeard SO user comes up with.}\\
\end{tabular}
\\
& 
Being patient 
&       
\begin{tabular}[t]{@{}l@{}l@{}}
\emph{"When you have a Chat bot, you have an infinitely patient college professor who will politely} \\
\emph{and endlessly answer any and all questions you have."}\\ 
\end{tabular}
\\
\bottomrule\bottomrule
\\

\end{longtable}
\end{scriptsize}

\begin{scriptsize}
\vspace{-0.1in}
\renewcommand\arraystretch{1.2}
\begin{longtable}{lll}
\caption{Example themes, sub-themes and quote examples that demonstrate the \textbf{Use Cases} of using ChatGPT to assist coding from Reddit posts.}
\label{tab:usercase}\\
\toprule\toprule
Themes                                                     & 
Sub-Themes                                                 & Quote Example \\ 
\midrule
 \begin{tabular}[t]{@{}l@{}}\textbf{Direct}\\ \textbf{Assistance}\end{tabular} 
& \begin{tabular}[t]{@{}l@{}}Seek guidance on\\ syntax/libraries/\\functions\end{tabular} 
& 
\begin{tabular}[t]{@{}l@{}l@{}}
\emph{"It is amazing for providing examples of syntax for new languages. I only just realized how much}\\
\emph{time I actually spend looking for that."}\\
\emph{"I used it today to ask how to use a 3rd party c\# library because the official documentation was}\\
\emph{lacking. It was helpful."}
\end{tabular}
\\

& \begin{tabular}[t]{@{}l@{}l@{}}Receive code\\ snippets/scripts\end{tabular}   
&  
\begin{tabular}[t]{@{}l@{}}
\emph{"It’s really great at generating isolated code snippets for solved problems and can be a real time}\\ 
\emph{saver in some, but certainly not all, cases."}
\end{tabular}
\\

& 
Debug                                                      
& 
\begin{tabular}[t]{@{}l@{}}
\emph{"I used chatGPT to help me debug and point out my logical mistakes and it's very helpful}\\\emph{regarding that."}
\end{tabular}
\\

& 
\begin{tabular}[t]{@{}l@{}l@{}}Generate\\ documentation\end{tabular} 
&    
\begin{tabular}[t]{@{}l@{}l@{}}
\emph{"It comments the code fairly well."}\\
\emph{"It can not only write code, it can also transpile, document, inline comment existing code... }
\end{tabular}
\\
\midrule
\begin{tabular}[t]{@{}l@{}}\textbf{Testing}\\ \textbf{Support}\end{tabular}  
& 
Generate unit tests                                                        
&  
\begin{tabular}[t]{@{}l@{}l@{}}
\emph{"It works for generating docstrings, unit tests, and example usage, given an untested}\\\emph{implementation."}
\end{tabular}
\\
 & 
 Manage edge cases                                                          
 &
 \begin{tabular}[t]{@{}l@{}}
 \emph{"It's ironic, ChatGPT has been able to solve all manner of weird and edge case code I've thrown at}\\
 \emph{it that would have taken a few hours to fully write and unit test otherwise."}
 \end{tabular}
 \\
\midrule
\begin{tabular}[t]{@{}l@{}}\textbf{Coding}\\\textbf{Solution}\end{tabular}                                                             & 
Seek start helper                                                  
&   
\emph{"I wouldn't have known where to even begin without this thing."}
\\
 & \begin{tabular}[t]{@{}l@{}}Get direction/\\ structure/outlet\end{tabular}  
 &  
 \begin{tabular}[t]{@{}l@{}}
 \emph{"It’s supposed to point you in a general direction and then you use something it doesn’t have: your}\\\emph{brain."}
 \end{tabular}
 \\
 &&\\
&
 \begin{tabular}[t]{@{}l@{}}Improve code/\\ efficiency\end{tabular}                                              
 &   
\begin{tabular}[t]{@{}l@{}}
 \emph{"I have been copy and pasting my code into ChatGPT and asking it to make it more efficient, and}\\
 \emph{after a few tries it comes up with some beautiful ideas."}
 \end{tabular}
 \\
 &&\\

 & \begin{tabular}[t]{@{}l@{}}Transit between\\ codes/languages\end{tabular} 
 & 
 \begin{tabular}[t]{@{}l@{}}
 \emph{"How do we take software built in 1980 using Fortran or assembly and quickly convert it to C\#,}\\
 \emph{Python, or Java while keeping current functionality in place and creating a seamless transition to}\\
 \emph{new applications and software."}
 \end{tabular}
 \\
\midrule
 \begin{tabular}[t]{@{}l@{}}\textbf{Provide }\\\textbf{Examples}\\ \textbf{and}\\\textbf{Explanations}\end{tabular} 
&                                                                            
&  
 \begin{tabular}[t]{@{}l@{}l@{}l@{}l@{}l@{}}
\emph{"I'm not great with JS so I pop it into ChatGPT and ask it to explain what the script is doing in}\\
\emph{detail, and it works beautifully."}\\
\emph{"I’ve struggled to teach myself to programming in the past because I have a hard time learning} \\
\emph{from just reading things on the internet. But with chatGPT I can have it show me example code,}\\
\emph{explain it, and answer any questions with more examples."}
 \end{tabular}
\\
\midrule

\begin{tabular}[t]{@{}l@{}}\textbf{Code}\\ \textbf{Assistant}\end{tabular}   
& 
\begin{tabular}[t]{@{}l@{}l@{}}Find pair\\ programming\\partner\end{tabular}           & 
\begin{tabular}[t]{@{}l@{}}
\emph{"I am a WebDev with 20yrs experience, and having an AI as a pair programming partner is the}
\\\emph{best thing that has happened to me in a long time in this industry. It won't replace us but augment}\\
\emph{our daily work. (GitHub co-pilot and ChatGPT)"}
\end{tabular}
\\
\midrule
\textbf{Training}                                  & 
Generate quiz                                                     
& 
\begin{tabular}[t]{@{}l@{}}
\emph{"Ask it questions like: "Can you give me a set of recursive problem exercises that I can try and solve}\\
\emph{on my own?" And it will reply with a couple of questions, along with the explanation if your lost."}
\end{tabular}
\\
\midrule
\textbf{Simulator}                                                                & Code editor                                                                
& 
 \begin{tabular}[t]{@{}l@{}l@{}}
\emph{"I'm trying to get it to work as a code editor and it does work, but I can't get it to stop giving me}\\
\emph{explanations. This is a fun text adventure game but even the big model is limited in how much state}\\
\emph{it can keep straight in its little context for you. So if you `mkdir` then do something else, it probably}\\
\emph{forgets the contents of its imaginary filesystem."}
 \end{tabular}
\\
\bottomrule\bottomrule
\\

\end{longtable}
\end{scriptsize}

\begin{scriptsize}
\renewcommand\arraystretch{1.2}
\begin{longtable}{lll}
\caption{Example themes, sub-themes and quote examples that demonstrate the \textbf{Barriers} of using ChatGPT to assist coding from Reddit posts.}
\label{tab:barriers}\\
\toprule\toprule
Themes &
  Sub-Themes &
  Quote Example \\ 
\midrule
\begin{tabular}[t]{@{}l@{}}
\textbf{Reliability} \\ \textbf{Challenges}\end{tabular} &
  Inconsistency in answers 
  &
  \begin{tabular}[t]{@{}l@{}}
  \emph{"It would make up different fake code every time it was asked the exact same"} \\
  \emph{"question."}
  \end{tabular}
\\
 &
  \begin{tabular}[t]{@{}l@{}}Fabrication of nonexistent \\ libraries/commands/citations\end{tabular} 
  &
\emph{"It can also make up classes and libraries to solve problems without telling you."}
   \\
   &&\\
 &
  \begin{tabular}[t]{@{}l@{}}Tone of response unaffected\\ by confidence level\end{tabular} 
  &
  \begin{tabular}[t]{@{}l@{}}
  \emph{"ChatGPT is absolutely excellent. But it is frequently wrong, and it's wrong with}\\
  \emph{calm and assured confidence."}
  \end{tabular} 
   \\
   &&\\
 &
  \begin{tabular}[t]{@{}l@{}}Unreliability when utilized\\ independently\end{tabular} &
  \begin{tabular}[t]{@{}l@{}}
  \emph{"Also, ChatGPT only helps if you can judge whether its answer is correct, unlike an}\\
 \emph{answer with 50 upvotes on Stack Overflow, which you \_know\_ is correct."}
  \end{tabular}
   \\
   \midrule
\begin{tabular}[t]{@{}l@{}}\textbf{Sensitivity to} \\ \textbf{Input}\end{tabular} &
  \begin{tabular}[t]{@{}l@{}}Responses influenced by \\ question phrasing\end{tabular} &
  \begin{tabular}[t]{@{}l@{}}
  \emph{"chatGPT now - I have to give it a super specific, context heavy explanation of what}\\
  \emph{I need from it. If I don't, it has a high probability of being wrong. "}
  \end{tabular}
   \\
   &&\\
 &
  Require clear/specific goals 
  &
  \begin{tabular}[t]{@{}l@{}l@{}l@{}}
 \emph{"I just tell it to write code that does a specific thing in a specific language and it}\\\emph{always works."}
   \end{tabular}
   \\
\midrule
\begin{tabular}[t]{@{}l@{}}\textbf{Lack of} \\ \textbf{Transparency}\end{tabular} &
  \begin{tabular}[t]{@{}l@{}}Lack of citations/links to \\ supporting documentation\end{tabular} 
  &
  \begin{tabular}[t]{@{}l@{}}
  \emph{"I just wish it could link to the relevant documentation."}\\
  \emph{"The content that ChatGPT creates is vanilla and without flair, no links,  no stats}\\\emph{or references."}
  \end{tabular}
   \\
 &
  \begin{tabular}[t]{@{}l@{}}Ambiguity surrounding \\ training sources\end{tabular} &
  \begin{tabular}[t]{@{}l@{}}
  \emph{"it lied to my face and said only on openais own sources. you can find the exact same}\\\emph{results on stack etc"}\\
  \emph{"Can someone point to the data and training requirements for ChatGPT?"}
  \end{tabular}
   \\
\midrule
   \begin{tabular}[t]{@{}l@{}}\textbf{Potential Code}\\\textbf{Management} \\ \textbf{Issues}\end{tabular} &
  \begin{tabular}[t]{@{}l@{}}Reduced code \\ maintainability\end{tabular} 
  &
  \begin{tabular}[t]{@{}l@{}}
  \emph{"AI will never be perfect enough to replace software developers, and will likely create}\\
  \emph{code that is not maintainable."}
  \end{tabular} 
   \\
 &
  \begin{tabular}[t]{@{}l@{}}Disruption of coding \\ style\end{tabular} 
  &
  \begin{tabular}[t]{@{}l@{}l@{}}
  \emph{"Please make sure the code it spits out matches with the existing coding architecture}\\
  \emph{of the project and is consistent, ...Its different coding style all over, it works but it}\\
  \emph{looks like someone coded it with split personality."}
  \end{tabular} 
   \\
\midrule
\begin{tabular}[t]{@{}l@{}}\textbf{Limited} \\ \textbf{Resource}\end{tabular} &
  Constrained training datasets &
  \begin{tabular}[t]{@{}l@{}}
  \emph{"It just occurred to me that AI faces the same challenge with generating code as it}\\ 
  \emph{does with anything else: it’s only as smart as its training set."}
  \end{tabular}
   \\
   \begin{tabular}[t]{@{}l@{}}\textbf{Access}\end{tabular} 
 &
  \begin{tabular}[t]{@{}l@{}}Absence of updated/online/\\ live resources\end{tabular} 
  &
  \begin{tabular}[t]{@{}l@{}}
  \emph{"ChatGPT’s LLM doesn’t include data beyond 2021, and most of the code I’m writing}\\
  \emph{has updated libraries/modules/packages that ChatGPT has no knowledge of. "}
  \end{tabular}
   \\
 &
   &
   \\
 &
  Inability to run locally &
  \begin{tabular}[t]{@{}l@{}}
  \emph{"Auto-Gpt uses OpenAI for inferencing but I want to use my local 13B alpaca instead.} \\
 \emph{Goal is 100\% local run and only go net when heavy lifts are needed."}
  \end{tabular}
   \\
 &
  Lack of IDE plugins 
  &
  \emph{"Unfortunately it doesn't have an IDE plugin for the IDE i use."}
   \\
\midrule
\begin{tabular}[t]{@{}l@{}}\textbf{Feature} \\ \textbf{Limitations}\end{tabular} &
  \begin{tabular}[t]{@{}l@{}}Refusal to address specific \\ queries\end{tabular} 
  &
  \begin{tabular}[t]{@{}l@{}}
 \emph{ "Try not to use the word "script". Whenever I do, it either comes up with some half} \\
  \emph{assed "sorry ai dumb, can't", or it will straight up just block up completely with an}\\\emph{error."}
    \end{tabular}
   \\
 &
  Features cut-off &
   \begin{tabular}[t]{@{}l@{}}
   \emph{"Yes recently it stop writing code"}
   \emph{"It was coding so well until they updated it and}\\
   \emph{killed it. Now it just makes up excuses or gets lost."}
     \end{tabular}
   \\
 &
  \begin{tabular}[t]{@{}l@{}}Absence of open-source/\\ customized versions for \\ adult users\end{tabular} 
  & 
  \begin{tabular}[t]{@{}l@{}l@{}}
  \emph{"We need an open source version. Something that doesn't treat us all like sensitive}\\
  \emph{children."}
  \end{tabular}
   \\
   \midrule
\begin{tabular}[t]{@{}l@{}}\textbf{Accessibility}\\\textbf{Constraints}\end{tabular} &
  Rate limitations 
  &
  \begin{tabular}[t]{@{}l@{}l@{}}
  \emph{"It's crazy how fast this could let people go if the API wasn't slow and rate limited}\\\emph{like it is now."}
  \end{tabular}
  \\

 &
  Regional restrictions &
  \begin{tabular}[t]{@{}l@{}l@{}}
  \emph{"OpenAI is not available for everyone, it has regional blocking, and if you use VPN,}\\
  \emph{they ask you to provide a valid phone number to continue."}
  \end{tabular}
   \\
 &
  Financial implications 
  &
  \begin{tabular}[t]{@{}l@{}l@{}l@{}}
  \emph{"They are gonna slice the AI up into pieces and charge \$20/month per skill... This will} \\
 \emph{ensure that nobody who is uneducated and poor can create something amazing."}
  \end{tabular}
   \\
 &
  Downtime/System overload &
 \emph{ "Didn't ChatGPt shutdown due to excessive usage?"}
   \\
\midrule
\begin{tabular}[t]{@{}l@{}}\textbf{Copyright} \\ \textbf{Concerns}\end{tabular} &
  \begin{tabular}[t]{@{}l@{}}Usage within professional \\ settings\end{tabular} &
  \begin{tabular}[t]{@{}l@{}}
  \emph{"I'm worried about copyright, trade secrets and NDAs so much that I would never}\\
  \emph{upload code that I write for my employer into ChatGPT"}
  \end{tabular}
   \\
 \bottomrule\bottomrule

\end{longtable}
\end{scriptsize}

\end{document}